\documentstyle[epsfig]{article}

\begin{document}

\title{NEW CONSTRAINTS ON WIMPS FROM THE CANFRANC
IGEX DARK MATTER SEARCH}

\author{}

\date{}

\maketitle

\begin{center}

A. Morales$^{a}$,
C.E. Aalseth$^{b}$,
F.T. Avignone III$^{b}$,
R.L. Brodzinski$^{c}$,
S. Cebri\'{a}n$^{a}$,
E. Garc\'{\i}a$^{a}$, \\
D. Gonz\'{a}lez$^{a}$,
W.K. Hensley$^{c}$,
I.G. Irastorza$^{a}$,
I.V. Kirpichnikov$^{d}$,
A.A. Klimenko$^{e}$,
H.S. Miley$^{c}$, \\
J. Morales$^{a}$,
A. Ortiz de Sol\'{o}rzano$^{a}$,
S.B. Osetrov$^{e}$,
V.S. Pogosov$^{f}$,
J. Puimed\'{o}n$^{a}$,
J.H.\ Reeves$^{c}$, \\
M.L. Sarsa$^{a}$,
S. Scopel$^{a}$,
A.A. Smolnikov$^{e}$,
A.G. Tamanyan$^{f}$,
A.A. Vasenko$^{e}$,
S.I. Vasiliev$^{e}$,
J.A. Villar$^{a}$
\end{center}

\begin{center}
\begin{em}

$^{a}$Laboratory of Nuclear and High Energy Physics, University of
Zaragoza, 50009 Zaragoza, Spain
\\
$^{b}$University of South Carolina, Columbia, South Carolina 29208
USA
\\
$^{c}$Pacific Northwest National Laboratory, Richland, Washington
99352 USA
\\
$^{d}$Institute for Theoretical and Experimental Physics, 117 259
Moscow, Russia
\\
$^{e}$Institute for Nuclear Research, Baksan Neutrino Observatory,
361 609 Neutrino, Russia
\\
$^{f}$Yerevan Physical Institute, 375 036 Yerevan, Armenia \\

\end{em}
\end{center}


\abstract{The IGEX Collaboration enriched $^{76}$Ge
double-beta decay detectors are currently operating in the
Canfranc Underground Laboratory with an overburden of 2450 m.w.e.
A recent upgrade has made it possible to use them in a search
for WIMPs.
A new exclusion plot,
$\sigma$(m), has been derived for WIMP-nucleon spin-independent
interaction. To obtain this result, 30 days of data from one IGEX
detector, which has an energy threshold $E_{thr} \sim 4$ keV, have
been considered.
These data improve the exclusion
limits
derived from other germanium diode experiments in the $\sim$50~GeV
DAMA region, and show that with a moderate improvement of the background
below 10 keV, the DAMA region may be tested
with an additional 1~kg-year of exposure.}

\section{Introduction}
Substantial evidence exists suggesting most matter in
the universe is dark, and there are compelling reasons to believe
it consists mainly of non-baryonic particles. Among these candidates,
Weakly Interacting
Massive and neutral Particles (WIMPs) are among the
front runners. The lightest stable
particles of supersymmetric theories, like the neutralino, describe a
particular
class of WIMPs\cite{Gri}.

Direct detection techniques rely on measurement of
WIMP elastic scattering off target nuclei in a
suitable detector\cite{Mor99}. Slow
moving ($\sim300$ km/s) and heavy ($10 - 10^3$ GeV)
galactic halo WIMPs could make a Ge nucleus recoil with a few keV, at
a rate which
depends on the type of WIMP and interaction. Only about 1/4 of this
energy is visible in the detector. Because of the low
interaction
rate and the small energy deposition, the direct search for particle
dark matter through scattering by
nuclear targets requires ultralow background detectors with very
low energy thresholds.

To detect the possible presence of WIMPs, the predicted event rate is
compared with the observed spectrum. If this predicted event rate
is larger than the measured one, the particle under
consideration can be ruled out as a dark matter component. Such
absence of WIMPs can be expressed as a contour line, $\sigma$(m), on
the WIMP-nucleus elastic scattering cross-section plane. This excludes,
for each mass, those particles whose cross-section lies above the contour
line, $\sigma$(m).

This direct comparison of the expected signal with the observed
background
spectrum can only exclude
or constrain the cross-section in terms of exclusion plots of $\sigma$(m).
A convincing proof of the detection of dark matter would require finding
unique signatures in the data, characteristic of the WIMP, which cannot
be attributed to the background or instrumental artifacts. An example is
the predicted summer-winter asymmetry \cite{Dru86} in the WIMP
signal rate due to the periodicity of the relative Earth-halo motion
resulting from the Earth's rotation around the Sun.

Germanium detectors used for double-beta decay searches have reached
one of the lowest background levels of any type of detector and  have a
reasonable quenching factor ($\sim 0.25$). Thus, with sufficiently low
energy thresholds, they are attractive devices for dark matter searches.

Germanium diodes dedicated to double-beta decay
experiments[4-11]
were applied to WIMP searches as early as
1987. The exclusion contour based on the best combination of data from
these experiments is referred to in this
paper as the
``combined germanium contour''.
Only recently has this exclusion plot been surpassed
by a sodium iodide experiment\cite{Ber96} (DAMA NaI-0),
which uses a statistical pulse-shape discriminated background spectrum.

This paper presents a new germanium detector data limit for the
direct detection of
non-baryonic particle dark matter in the $\sim$50~GeV DAMA mass
region.

\section{Experiment}
The IGEX experiment\cite{Aal,Gon99}, optimized for detecting
$^{76}$Ge double-beta
decay, has been described in detail elsewhere. The IGEX
detectors are now also being used in the
search for WIMPs interacting coherently with germanium nuclei. The
COSME detector described below, is also
operating in the same shield at Canfranc.

The IGEX detectors were fabricated at Oxford
Instruments, Inc., in Oak Ridge, Tennessee.
Russian GeO$_2$ powder, isotopically enriched to 86\% $^{76}$Ge,
was purified, reduced to metal, and zone
refined to $\sim 10^{13}$ p-type donor impurities per cubic centimeter
by Eagle Picher, Inc., in Quapaw, Oklahoma. The metal was then
transported to Oxford Instruments by surface in order to minimize
activation by cosmic ray neutrons, where it was further zone refined,
grown into crystals, and fabricated into detectors.

The COSME detector was fabricated at Princeton Gamma-Tech, Inc. in
Princeton, New Jersey, using naturally abundant germanium. The
refinement of newly-mined germanium ore to finished metal for this
detector was expedited to minimize production of cosmogenic
$^{68}$Ge.

All of the cryostat parts were electroformed using a high purity OFHC
copper/CuSO$_4$/H$_2$SO$_4$ plating system.
The solution was continuously filtered to eliminate copper
oxide, which causes porosity in the copper. A Ba(OH)$_2$ solution  was
added to precipitate BaSO$_4$, which is also collected
on the filter. Radium in the bath exchanges with the barium on the filter,
thus minimizing radium contamination
in the cryostat parts. The CuSO$_4$ crystals were
purified of thorium by
multiple recrystallization.

The IGEX detector used for dark matter searches, designated RG-II, has
a mass of $\sim2.2$ kg. The active mass of this detector, $\sim2.0$~kg,
was measured with a collimated source of $^{152}$Eu in the Canfranc
Laboratory and is in agreement with the Oxford Instruments efficiency
measurements. The full-width at half-maximum (FWHM) energy
resolution of RG-II was 2.37~keV at the 1333-keV line of $^{60}$Co.
The COSME detector has a mass of 254~g and an active mass of 234~g.
The FWHM energy resolution of COSME is 0.43~keV at the
10.37~keV gallium X-ray. Energy calibration and resolution
measurements were made every 7--10 days using the lines of
$^{22}$Na and $^{60}$Co. Calibration for the low energy region was
extrapolated using the X-ray lines of Pb.

For each detector, the first-stage field-effect transistor (FET) is
mounted
on a Teflon block a few centimeters from the center contact of the
germanium crystal. The protective cover of the FET and the glass shell
of the feedback resistor have been removed to reduce radioactive
background. This first-stage assembly is mounted behind a 2.5-cm-thick
cylinder of archaeological lead to further reduce background. Further
stages of preamplification are located at the back of the cryostat cross
arm, approximately 70 cm from the crystal. The IGEX detectors have
preamplifiers modified for the pulse-shape analysis used in the
double-beta
decay searches.

The detectors shielding is as follows,
from inside to outside. The innermost shield consists of 2.5
tons of 2000-year-old archaeological lead forming a 60-cm cube and
having $<9$~mBq/kg of
$^{210}$Pb($^{210}$Bi), $< 0.2$~mBq/kg of $^{238}$U,
and $<0.3$~mBq/kg of $^{232}$Th.
The detectors fit into precision-machined holes in this central core,
which minimizes the empty space around the detectors available to
radon. Nitrogen gas, at a rate of 140~l/hour, evaporating from
liquid nitrogen, is forced into the detector chambers to create
a positive pressure and further minimize radon intrusion. The
archaeological lead block is centered in a 1-m cube of
70-year-old
low-activity lead($\sim 10$ tons) having $\sim 30$~Bq/kg of
$^{210}$Pb.
A minimum of 15~cm of archaeological lead separates the detectors
from
the outer lead shield.
A 2-mm-thick cadmium sheet surrounds the main lead shield, and two
layers of plastic seal this central assembly against radon intrusion. A
cosmic muon veto covers the top and sides of the central core, except
where the detector Dewars are located. The veto consists of BICRON
BC-408 plastic scintillators 5.08 cm $\times$ 50.8 cm $\times$ 101.6 cm
with surfaces finished
by diamond mill to optimize internal reflection. BC-800 (UVT) light
guides on the ends taper to 5.08 cm in diameter over a length of
50.8 cm and are coupled to Hamamatsu R329 photomultiplier tubes.
The anticoincidence veto signal is obtained from the logical OR of
all photomultiplier tube discriminator outputs.
An external polyethylene neutron moderator 20~cm thick (1.5 tons)
completes the shield. The entire shield is supported by an iron structure
resting on noise-isolation blocks. The experiment is located in a
room isolated from the rest of the laboratory and has an overburden of
2450 m.w.e., which reduces the measured muon flux to
$2 \times 10^{-7} \rm cm^{-2} \rm s^{-1}$.

The data acquisition system for the low-energy region used in dark
matter searches (referred to as IGEX-DM) is based on
standard NIM electronics and is independent from that used for
double-beta decay searches (IGEX-$2\beta$). It has been implemented
by splitting the normal preamplifier output pulses of each detector
and routing them through two Canberra 2020 amplifiers having different
shaping times enabling noise rejection\cite{Jmor}. These amplifier
outputs are
converted using 200 MHz Wilkinson-type
Canberra analog-to-digital converters, controlled by a PC through
parallel interfaces. For each event, the arrival time (with an accuracy of
100~$\mu$s), the elapsed time
since the last veto event (with an accuracy of 20~$\mu$s), and the
energy from each ADC are
recorded.

\section{Results}
The IGEX-DM results obtained correspond to 30 days of analyzed data
(Mt=60 kg-days) from IGEX detector RG-II. Also presented for
comparison are earlier
results from the COSME detector (COSME-1) \cite{Jmor,Gar92}, as well as recent
results obtained in its current set-up (COSME-2)\cite{Ceb00,Mor99}.

\begin{figure}[ht]
\centerline{
\epsfxsize=10cm
\epsffile{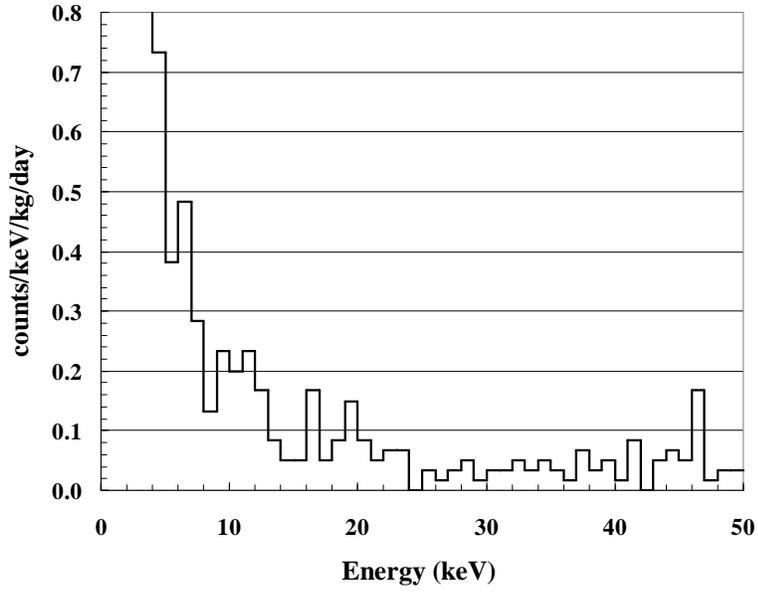}
}
 \caption{Low-energy spectrum of the IGEX RG-II detector
(Mt $=$ 60~kg-d).}
 \label{dm-ig-1}
\end{figure}

The detector RG-II features an energy threshold of 4 keV and an energy
resolution of
0.8 keV at the 75~keV Pb x-ray line. The background rate
recorded was $\sim 0.3$ c/(keV-kg-day) between 4--10~keV,
$\sim 0.07$ c/(keV-kg-day) between 10--20~keV, and
$\sim 0.05$ c/(keV-kg-day) between 20--40~keV. Fig.~\ref{dm-ig-1}
shows the RG-II 30-day spectrum; the numerical data are given in
Table~\ref{tab-ig-1}.

\begin{figure}[ht]
\centerline{
\epsfxsize=10cm
\epsffile{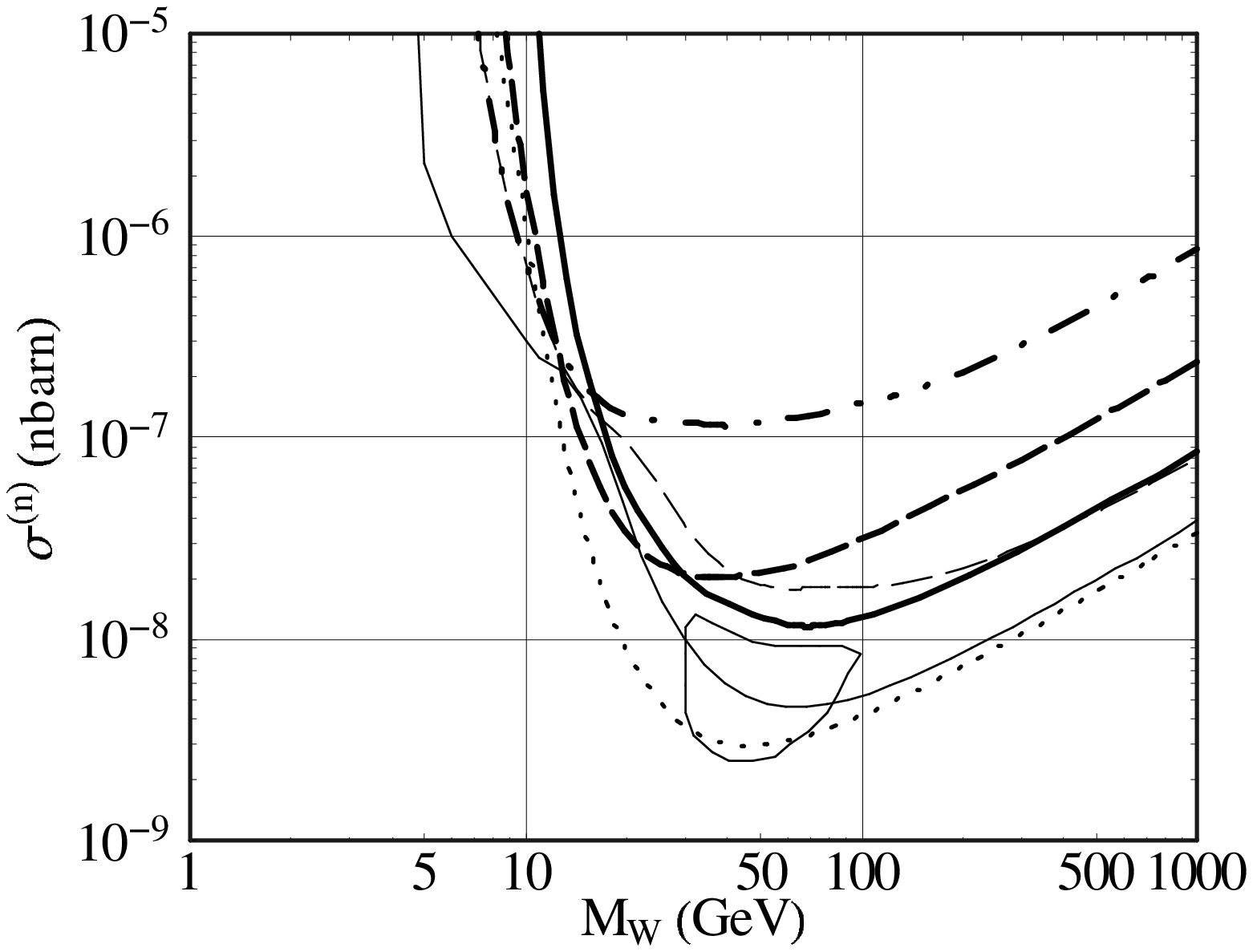}
}
 \caption{IGEX-DM exclusion plot for spin-independent
interaction obtained in this work
(thick solid line). Results obtained
in other Germanium experiments are also shown: Canfranc COSME-1 data\cite{Gar92}
(dot-dashed line), recent COSME-2 data\cite{Mor99,Ceb00} (thick
dashed line), and the previous Ge-combined bound (thin dashed line)
---including the last
Heidelberg-Moscow data\cite{Bau}. The result of the DAMA NaI-0 experiment
\cite{Ber96} (thin solid line) is also shown. The "triangle" area
corresponds to the (3$\sigma$) annual modulation effect reported by
the DAMA collaboration (including NaI-1,2,3,4 runnings)\cite{Ber99}.
The IGEX-DM projection (dotted line) is shown for
1~kg-year of exposure
 with a background rate of 0.1~c/(keV-kg-day).}
 \label{dm-ig-2}
\end{figure}

The exclusion plots are derived from the recorded spectrum in one-keV
bins from 4~keV to 50~keV.
As recommended by the Particle Data Group, the predicted signal in an
energy bin is required to be less than or equal to the (90\% C.L.) upper
limit of the (Poisson) recorded counts.
The derivation of the interaction rate signal supposes that the WIMPs
form an isotropic, isothermal, non-rotating halo of density
$\rho = 0.3$~GeV/cm$^{3}$, have a Maxwellian velocity distribution
with $\rm v_{\rm rms}=270$~km/s (with an upper cut corresponding to
an escape velocity of 650~km/s), and have a relative Earth-halo velocity
of $\rm v_{\rm r}=230$~km/s. The cross sections are normalized to the
nucleon, assuming a dominant
scalar interaction. The Helm parameterization\cite{Eng91} is used for
the scalar nucleon form factor, and the quenching factor used is 0.25.
The exclusion plots derived from the IGEX-DM (RG-II) and COSME
data are shown in Fig.~\ref{dm-ig-2}. In particular, IGEX results exclude
WIMP-nucleon cross-sections above 1.3x10$^{-8}$ nb for masses corresponding
to the 50 GeV DAMA region\cite{Ber99}. Also shown is the combined
germanium contour, including the last
Heidelberg-Moscow data\cite{Bau} (recalculated from the original
energy spectra with the same set of hypotheses and parameters), the
DAMA experiment contour plot derived from Pulse Shape
Discriminated spectra\cite{Ber96}, and the DAMA region corresponding to their
reported annual modulation effect\cite{Ber99}. The IGEX-DM
exclusion contour improves significantly on that of other germanium
experiments for masses corresponding to that of
the neutralino tentatively assigned to the DAMA modulation
effect\cite{Ber99} and results from using only unmanipulated data.

Data collection is currently in progress with improved background
below 20~keV.
Based on present IGEX-DM performance and reduction of the
background to $\sim 0.1$~c/(keV-kg-day) between 4--10~keV, the
complete DAMA region (m=$52^{+10}_{-8}$ GeV,
$\sigma$$^p$=($7.2^{+0.4}_{-0.9}$)x10$^{-9}$ nb)
could be tested after an exposure of 1~kg-year,
i.e. a few months of operation with two upgraded IGEX detectors.

\section*{Acknowledgements}
The Canfranc Astroparticle Underground Laboratory is operated by
the University of Zaragoza under contract No. AEN99-1033. This
research was partially funded by the Spanish Commission for
Science and Technology (CICYT), the U.S. National Science
Foundation, and the U.S. Department of Energy. The isotopically
enriched $^{76}$Ge was supplied by the Institute for Nuclear
Research (INR), Moscow, and the Institute for Theoretical and
Experimental Physics (ITEP), Moscow.

\newpage

\begin{table}[htb]
\begin{center}
\begin{tabular}[h]{cccccc}
\hline \multicolumn{1}{r}{{\bf E (keV)}}
                 & \multicolumn{1}{r}{{\bf counts}}

                 & \multicolumn{1}{r}{{\bf E (keV)}}
                 & \multicolumn{1}{r}{{\bf counts}}

                 & \multicolumn{1}{r}{{\bf E (keV)}}
                 & \multicolumn{1}{r}{{\bf counts}}
                  \\
\hline
\small
4.5 & 44 & 19.5 & 9 & 34.5 & 3\\ 5.5 & 23 & 20.5 & 5 & 35.5 & 2\\
6.5 & 29 & 21.5 & 3 & 36.5 & 1\\ 7.5 & 17 & 22.5 & 4 & 37.5 & 4\\
8.5 & 8 & 23.5 & 4 & 38.5 & 2\\ 9.5 & 14 & 24.5 & 0 & 39.5 & 3\\
10.5 & 12 & 25.5 & 2 & 40.5 & 1\\ 11.5 & 14 & 26.5 & 1 & 41.5 &
5\\ 12.5 & 10 & 27.5 & 2 & 42.5 & 0\\ 13.5 & 5 & 28.5 & 3 & 43.5 &
3\\ 14.5 & 3 & 29.5 & 1 & 44.5 & 4\\ 15.5 & 3 & 30.5 & 2 & 45.5 &
3\\ 16.5 & 10 & 31.5 & 2 & 46.5 & 10\\ 17.5 & 3 & 32.5 & 3 & 47.5
& 1\\ 18.5 & 5 & 33.5 & 2 & 48.5 & 2\\ \hline
\end{tabular}
\caption{Low-energy data from the IGEX RG-II detector
(Mt $=$ 60~kg-d).}
\label{tab-ig-1}
\end{center}
\end{table}

\newpage

\newpage

\begin{thebibliography}{99}
\bibitem{Gri} G. Jungman, M. Kamionkowski and K. Griest, Phys.
Rep. 267 (1996) 195
\bibitem{Mor99} For a recent survey of WIMP detection, see for
instance A. Morales ``Direct Detection of WIMP Dark
Matter'' (astro-ph/9912554). Review Talk at the TAUP 99 Workshop,
College de France, Paris (September 1999). To be published in Nucl.
Phys. B (Proc.
Suppl.) (2000)
\bibitem{Dru86} A.K. Drukier et al., Phys. Rev. D33 (1986) 3495
\bibitem{Ahl87} S.P. Ahlen et al., Phys. Lett. B195 (1987) 603
\bibitem{Cal88} D.O. Caldwell et al., Phys. Rev. Lett. 61 (1988) 510
\bibitem{Reu91} D. Reusser et al., Phys. Lett. B255 (1991) 143
\bibitem{Jmor} J. Morales et al., Nucl. Instrum. Meth. A321 (1992)
410
\bibitem{Gar92} E. Garc\'{\i}a et. al., Nucl. Phys. B (Proc. Suppl.)
28A (1992)286, Phys. Rev. D51 (1995) 1428
\bibitem{Dru92} A.K. Drukier et al., Nucl. Phys. B (Proc. Suppl.)
28A (1992) 293
\bibitem{Bec94} M. Beck et al., Phys. Lett. B336 (1994) 141
\bibitem{Bau} L. Baudis et al., Phys. Rev. D59 (1998) 022001
\bibitem{Ber96} R. Bernabei et al., Phys. Lett. B379 (1996) 299
\bibitem{Aal} C. Aalseth et al., Phys. Rev.  C59 (1999) 2108
\bibitem{Gon99} D. Gonz\'{a}lez et al., ``Current IGEX results for
neutrinoless
double beta decay of Ge-76'', talk given at the TAUP 99 Workshop,
College de France, Paris (September 1999). To be published in Nucl.
Phys. B
(Proc. Suppl.) 2000
\bibitem{Ceb00} S. Cebri\'{a}n et al., to appear in New Journal of
Physics, (2000)
\bibitem{Eng91} J. Engel, Phys. Lett. B264 (1991) 114
\bibitem{Ber99} R. Bernabei et al., Phys. Lett. B450 (1999) 448,
ROM2F/2000/01, January 2000
\end{thebibliography}
\end{document}